\documentclass[12pt]{article}

\usepackage{sbc-template}
\usepackage{graphicx,url}
\usepackage[utf8]{inputenc}
\usepackage[english]{babel}
\usepackage{subcaption}
\usepackage{amsmath} 
\usepackage{float}
\usepackage{acronym}
\usepackage{xcolor}
\usepackage{fancybox}
\usepackage{enumitem}

\usepackage[absolute]{textpos}
\usepackage{tikz}

\title{Inspiring Women in Technology: Educational Pathways and Impact}
\author{Larissa F. {Rodrigues Moreira}\inst{1}, Liziane S. Soares\inst{1},
Adriana Z. Martinhago\inst{1}}
\address{Institute of Exacts and Technological Sciences -- Federal University of Viçosa (UFV)\\
  Rio Paranaíba -- MG -- Brazil
  \email{\{larissa.f.rodrigues, adriana.martinhago, liziane.soares\}@ufv.br}
}

\begin{document} 
%Numbering
\acrodef{3GPP}{3rd Generation Partnership Project}
%-----A-----
\acrodef{AI}{Artificial Intelligence}
%-----B-----
\acrodef{B5G}{Beyond Fifth Generation}

%-----C-----
\acrodef{CUBIC}{Conjunctive Using BIC (Binary Increase Congestion Control)}
\acrodef{cwnd}{Congestion Window}
%-----D-----
\acrodef{DoS}{Denial of Service}
\acrodef{DDoS}{Distributed Denial of Service}
\acrodef{DNN}{Deep Neural Network}
\acrodef{DRL}{Deep Reinforcement Learning}
\acrodef{DT}{Decision Tree}
\acrodef{DNN}{Deep Neural Network}
\acrodef{DMP} {Deep Multilayer Perceptron}
\acrodef{DQN}{Deep Q-Learning}
%-----E-----
\acrodef{ETSI}{European Telecommunications Standards Institute}
%-----F-----
\acrodef{FIBRE}{Future Internet Brazilian Environment for Experimentation}
\acrodef{FTP}{File Transfer Protocol}
\acrodef{Flat}{Flat Neural Network}
%-----G-----
\acrodef{GNN}{Graph Neural Networks}
%-----H-----
\acrodef{HTM}{Hierarchical Temporal Memory}

%-----I-----
\acrodef{IAM}{Identity And Access Management}
\acrodef{IID}{Informally, Identically Distributed}
\acrodef{IoE}{Internet of Everything}
\acrodef{IoT}{Internet of Things}
%-----J-----
%-----K-----
\acrodef{KNN}{K-Nearest Neighbors}
%-----L-----
\acrodef{LSTM}{Long Short-Term Memory}
%-----M-----
\acrodef{MPTCP}{Multipath Transmission Control Protocol}
\acrodef{M2M}{Machine to Machine}
\acrodef{MAE}{Mean Absolute Error}
\acrodef{ML}{Machine Learning}
\acrodef{MOS}{Mean Opinion Score}
\acrodef{MAPE}{Mean Absolute Percentage Error}
\acrodef{MSE}{Mean Squared Error}
\acrodef{mMTC}{Massive Machine Type Communications}
\acrodef{MFA}{Multi-factor Authentication}
\acrodef{MQTT}{Message Queuing Telemetry Transport}

%-----N-----
\acrodef{NN}{Deep Neural Network}
\acrodef{NS3}{Network Simulator 3}
%-----O-----
\acrodef{OSM}{Open Source MANO}
%-----P-----
%-----Q-----
\acrodef{QL}{Q-learning}
\acrodef{QoE}{Quality of experience}
\acrodef{QoS}{Quality of Service}
%-----R-----
\acrodef{RAM}{Random-Access Memory}
\acrodef{RF}{Random Forest}
\acrodef{RL}{Reinforcement Learning}
\acrodef{RMSE}{Root Mean Square Error}
\acrodef{RNN}{Recurrent Neural Network}
\acrodef{Reno}{Regular NewReno}
\acrodef{RTT}{Round Trip Time}
%-----S-----
\acrodef{SDN}{Software-Defined Networking}
\acrodef{SFI2}{Slicing Future Internet Infrastructures}
\acrodef{SLA}{Service-Level Agreement}
\acrodef{SON}{Self-Organizing Network}

%-----T-----
\acrodef{TCP}{Transmission Control Protocol}
%-----U-----
%-----V-----
\acrodef{VoD}{Video on Demand}
\acrodef{VR}{Virtual Reality}
\acrodef{V2X}{Vehicle-to-Everything}

%-----W-----
%-----X-----
%-----Y-----
%-----Z-----
% Configura o ambiente para posicionamento absoluto
\begin{textblock*}{15cm}(3cm,27.8cm) % (Largura, (x,y))
\noindent
    \footnotesize\textcolor{red}{This paper has been accepted by the Women in Information Technology (WIT) 2024. 
    The definite version of this work was published by SBC-OpenLib as part of the WIT conference proceedings. 
    \\DOI: \url{https://doi.org/10.5753/wit.2024.1910}}
\end{textblock*}

\maketitle

\begin{abstract}
%This paper presents initiatives aimed at fostering female involvement in the realm of computing and endeavoring to inspire more women to pursue careers in these fields. The $<$omitido para revisão$>$ Project coordinates activities at both the secondary and higher education levels, facilitating dialogue between young women and computing professionals and promoting female role models within the field. Our findings indicate that XX\% of female high school students demonstrated interest in computing following their participation in the project's Ada Lovelace Day activities. Additionally, higher education initiatives have stimulated the engagement of both women and men, fostering inclusivity, entrepreneurship, and cooperation to improve the representation of women in the computing field.
This paper presents initiatives aimed at fostering female involvement in the realm of computing and endeavoring to inspire more women to pursue careers in these fields. The Meninas++ Project coordinates activities at both the high school and higher education levels, facilitating dialogue between young women and computing professionals, and promoting female role models within the field. Our study demonstrated the significant impact of these activities on inspiring, empowering, and retaining female students in computing. Furthermore, higher education initiatives have fostered engagement among both women and men, promoting inclusivity, entrepreneurship, and collaboration to enhance women’s representation in the computing field.
\end{abstract}

\section{Introduction}\label{sec:introduction}

Despite the progress made in promoting gender diversity, women continue to be underrepresented in the fields of Science, Technology, Engineering, and Mathematics (STEM)~\cite{Santos2023}.  Consequently, gender stereotypes, low female representation, and negative opinions of influential individuals within women's social circles contribute to the low presence of women in undergraduate STEM courses~\cite{Santos2023, Martinez2023}. 

To overcome this gap, diverse initiatives have been implemented worldwide and in Brazil~\cite{Barioni2022, Irion2023, Santos2023, Novaes2023}.  These initiatives aim to inspire and empower young women to enter STEM fields. They provide mentorship, hands-on learning opportunities, and access to successful female role models, thereby creating a supportive and encouraging environment for women~\cite{BolanFrigo2023}. 
%To overcome this gap, diverse initiatives have been implemented worldwide and in Brazil~\cite{Santos2023, Novaes2023}. 
In this sense, the Meninas++ Project~\cite{Nunes2015wie, Nunes2015} at the Federal University of Viçosa (UFV) disseminates knowledge about computing to undergraduate and high-school women, aiming to attract women to STEM and empower them with the skills and confidence necessary to thrive in the field. 
In this paper, we discuss our initiatives and their impacts on both high school and undergraduate students. The main contributions are: 

\begin{itemize}
    \item Provide a comprehensive analysis of Meninas++ Project initiatives to address the under-representation of women in STEM, particularly in computing.

    \item Demonstrating the positive impact of Meninas++ Project on inspiring, empowering, and retaining female students through quantitative data and qualitative feedback.

    \item Offering insights into the effectiveness of Meninas++ Project in fostering a supportive learning environment and equipping women with skills for success in computing careers.

    %\item Contributing with the field of gender diversity in STEM education for policymakers, educators, and stakeholders.

\end{itemize}

\section{Project Actions}\label{sec:actions}
\begin{figure}[!ht]
  \centering
  \includegraphics[width=\linewidth]{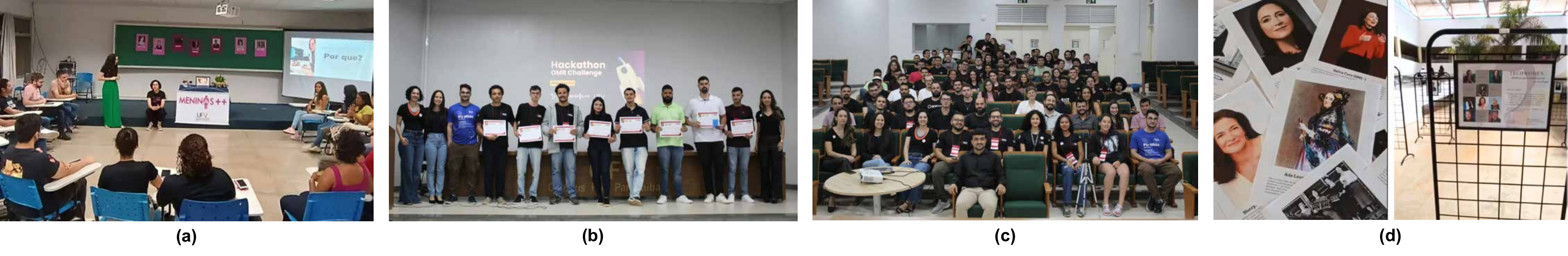}
  \caption{(a) Lectures and conversation groups, (b) Hackathon, (c) Workshop, and (d) TechWomen Exhibition.}
  \label{fig:events}
\end{figure}

We present in Table~\ref{tab:initiatives} an overview of the various initiatives undertaken by the Meninas++ Project to encourage women's participation and interest in technology. Figure~\ref{fig:events} shows pictures depicting diverse organized activities.
\begin{table}[!ht]
\centering
\caption{Summary of Meninas++ Project Actions.}
\scriptsize
\begin{tabular}{|p{2.5cm}|p{6cm}|p{5.3cm}|}
\hline
\multicolumn{1}{|c|}{\textbf{Initiative}} & \multicolumn{1}{c|}{\textbf{Description}} & \multicolumn{1}{c|}{\textbf{Goals}} \\ \hline

\textbf{Lectures} & 
- IT Market over 10 years, \newline - AI advancements, \newline - Ada Lovelace Day celebrations. & 
- Disseminate knowledge. \newline 
- Foster engagement. \newline 
- Inspire women in Computer. \\ \hline

\textbf{Exhibition: \newline TechWomen} & 
- Gallery at UFV highlighting women's contributions to computing history through images and narratives. & 
- Highlight women's achievements. \newline 
- Inspire and empower current and future generations. \\ \hline

\textbf{Conversation \newline Groups} & 
- Dialogue and collaboration among undergraduates for sharing insights and providing mutual support. & 
- Strengthen bonds. \newline 
- Offer mutual support. \newline 
- Help perseverance in studies. \\ \hline

\textbf{Meninas++ \newline Workshop} & 
- Annual event since 2017 with speakers sharing their journeys to inspire women to pursue tech careers. \newline 
- Networking and mentorship opportunities. \newline 
- Promote computing careers among high school students in Alto Paranaíba, MG. & 
- Motivate and empower participants. \newline 
- Foster networking and mentorship. \newline 
- Popularize computing. \\ \hline

\textbf{Hackathon OMR \newline Challenge} & 
- Collaboration with Pix Force to develop AI-based solutions for educational contexts. \newline 
- Participants created prototypes for automating multiple-choice test corrections using AI and Computer Vision. & 
- Promote innovation in technology. \newline 
- Encourage practical application of AI in education. \\ \hline
\end{tabular}
\label{tab:initiatives}
\end{table}

\section{Results and Discussion}\label{sec:results_and_discussion}

%In this section, we present the results of this study, including both quantitative data and qualitative insights, to provide a comprehensive analysis of the impact of the initiatives implemented by the Meninas++ Project.

\begin{figure}[!ht]
  \centering
  \begin{minipage}{0.3\textwidth}
    \centering
    \includegraphics[width=\linewidth]{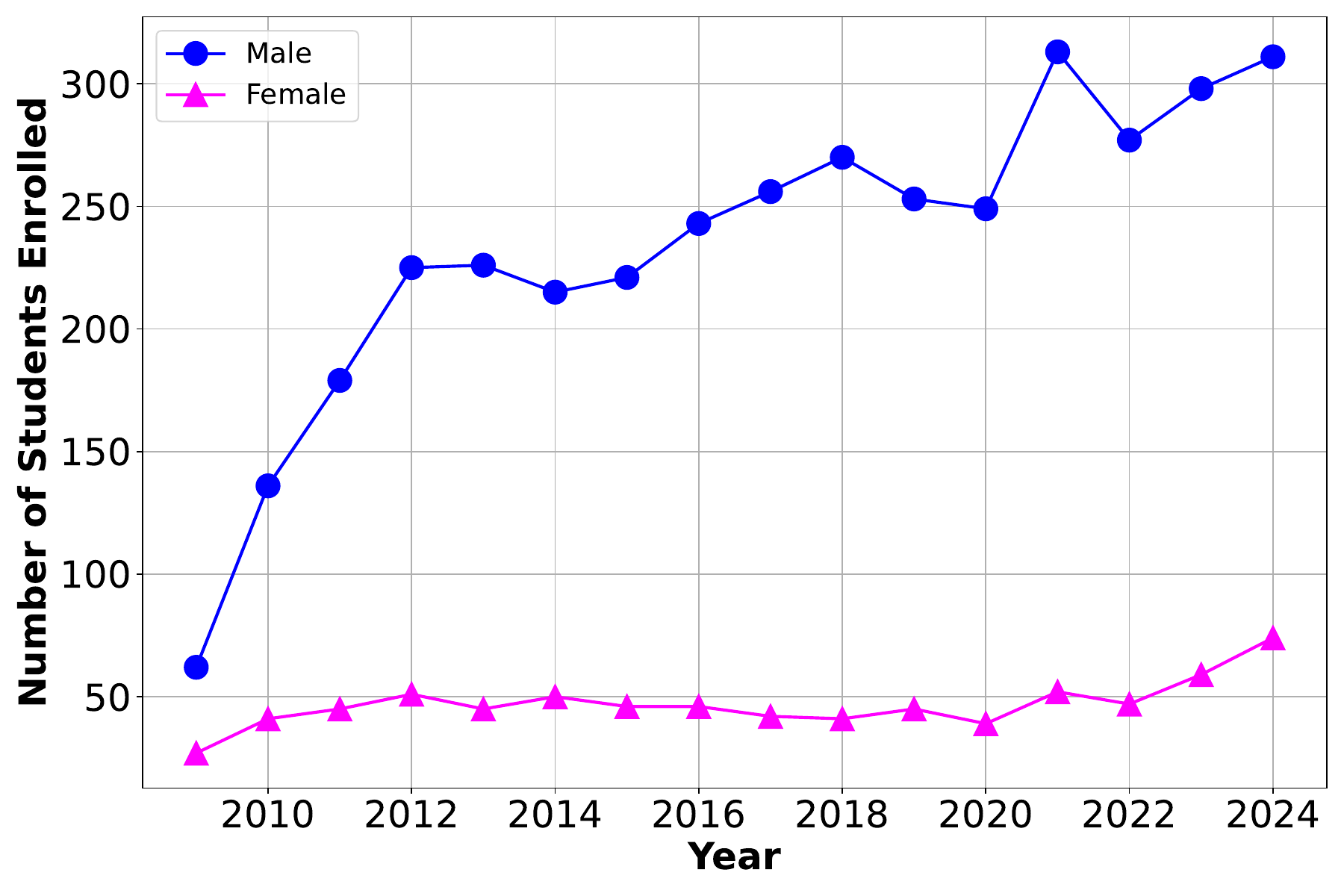}
    \caption{Enrollment per year and gender.}
    \label{fig:matriculados}
  \end{minipage}
  \hfill
  \begin{minipage}{0.3\textwidth}
    \centering
    \includegraphics[width=\linewidth]{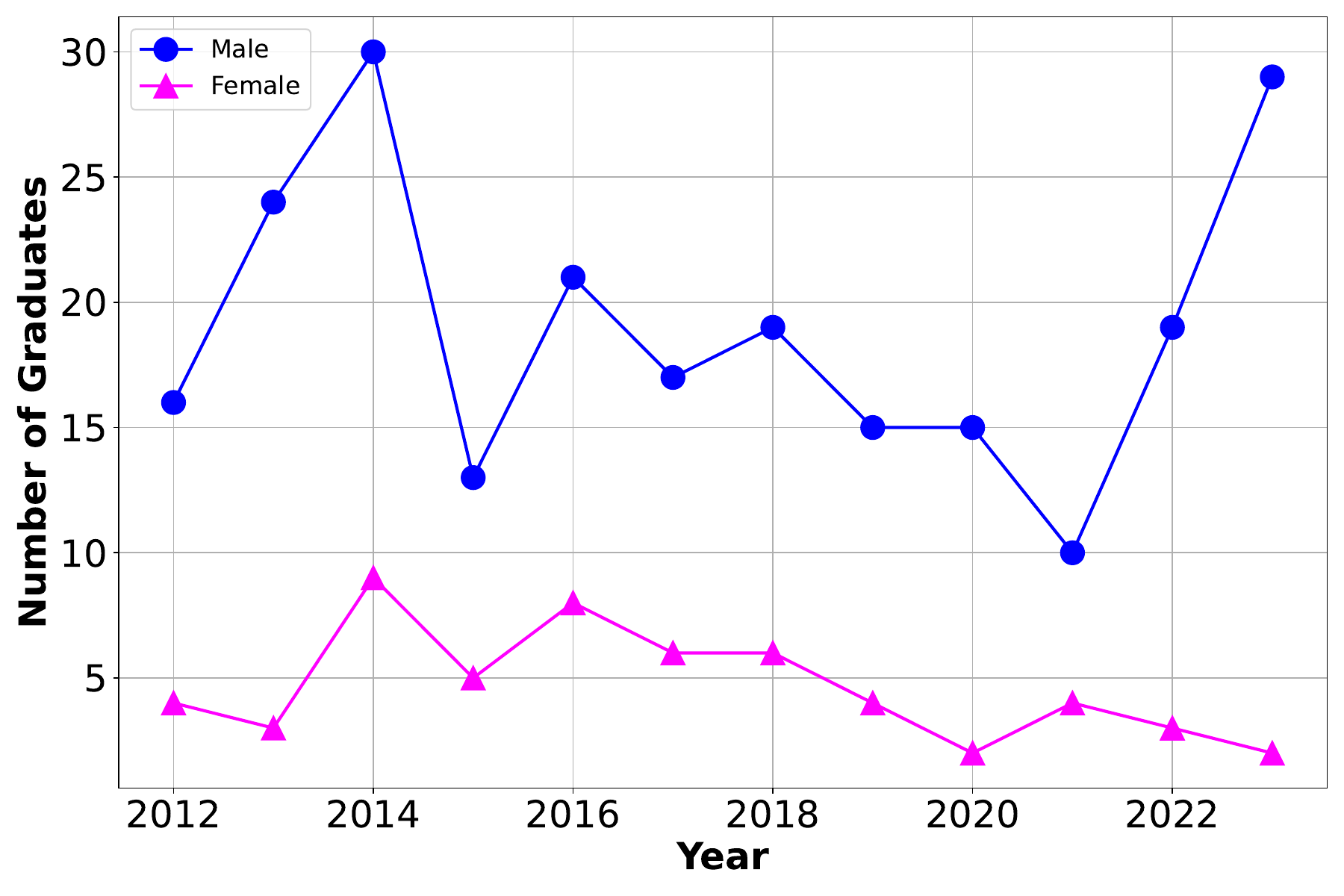}
    \caption{Graduates per year and gender.}
    \label{fig:formados}
  \end{minipage}
  \hfill
  \begin{minipage}{0.32\textwidth}
    \centering
    \tiny
    \captionof{table}{Entrance growth rate.} % Note o uso de \captionof{table} para a legenda da tabela
    \begin{tabular}{ccc}
    \hline
    \textbf{Year} & \textbf{Male (\%)} & \textbf{Female (\%)} \\ \hline
    2010          & 119.35             & 51.85                \\
    2011          & 31.62              & 9.76                 \\
    2012          & 25.70              & 13.33                \\
    2013          & 0.44               & -11.76               \\
    2014          & -4.87              & 11.11                \\
    2015          & 2.79               & -8.00                \\
    2016          & 9.95               & 0.00                 \\
    2017          & 5.35               & -8.70                \\
    2018          & 5.47               & -2.38                \\
    2019          & -6.30              & 9.76                 \\
    2020          & -1.58              & -13.33               \\
    2021          & 25.70              & 33.33                \\
    2022          & -11.51             & -9.62                \\
    2023          & 7.58               & 25.53                \\
    2024          & 4.36               & 25.42                \\ \hline
    \end{tabular}
    \label{tab:growth_rate}
  \end{minipage}
  %\caption{Number of regular and graduate students, and entrance growth rate.}
  \label{fig:figs_and_table}
\end{figure}

We analyzed the number of students enrolled and graduates of Information Systems courses at UFV, as these are the only computing courses available on the Rio Paranaíba Campus. On the university platform, we collected data that revealed a low representation of women in Information Systems courses, both as enrolling students (regular students) and graduates. Figure~\ref{fig:matriculados} depicts the enrollment rate of students in the undergraduate program under analysis, showing the enrollment of both male and female students. Figure~\ref{fig:formados} illustrates the graduation completion rates of these students over time. To assess the correlation between the enrollment rates of male and female students, we conducted a hypothesis test based on the following premises.

\begin{itemize}
    \item Null hypothesis ($\mathcal{H}_{0}$): There is no significant difference in the growth rate of enrollment between male and female students over time.

    \item Alternative hypothesis ($\mathcal{H}_{1}$): The growth rate of enrollment among male students is significantly higher than the growth rate of enrollment among female students over time.
\end{itemize}

%\begin{itemize}
   % \item Hipótese nula ($\mathcal{H}_{0}$): Não há diferença significativa na taxa de crescimento dos ingressos de estudantes masculinos e femininos ao longo do tempo.
    %\item Hipótese alternativa ($\mathcal{H}_{1}$): A taxa de crescimento dos ingressos de estudantes masculinos é significativamente maior do que a taxa de crescimento dos ingressos de estudantes femininos ao longo do tempo.
%\end{itemize}

%Considerando essa hipótese, calculamos a taxa de crescimento do ingresso ao longo dos anos conforme \text{Growth Rate} = $\frac{\text{Current Value} - \text{Previous Value}}{\text{Previous Value}} \times 100$. Calculamos a média e o desvio padrão das taxas de crescimento para cada grupo conforme Table~\ref{tab:growth_rate} e, em seguida, realizamos o teste t-Student para determinar se há uma diferença significativa entre as taxas de crescimento médias. As médias e desvios padrão das taxas de crescimento encontradas foram: Taxa de crescimento masculino: média = 2.09, desvio padrão = 12.39, Taxa de crescimento feminino: média = 2.77, desvio padrão = 15.45. Calculamos o valor crítico do teste t-Student para um nível de significância de 5\% e realizamos o teste para determinar se rejeitamos ou não a hipótese nula.
Considering this hypothesis, we calculate the enrollment Growth Rate over the years as \text{Growth Rate} = $\frac{\text{Current Value} - \text{Previous Value}}{\text{Previous Value}} \times 100$. We computed the mean and standard deviation of the growth rates for each group, as shown in Table~\ref{tab:growth_rate}, and then conducted the $T$-Student test to determine if there was a significant difference between the mean growth rates. The mean and standard deviation of the growth rates were as follows: male growth rate, mean $= 2.09$, standard deviation $= 12.39$; female growth rate, mean = $2.77$, standard deviation $= 15.45$. We calculated the critical value of the $T$-Student’s test at a significance level of 5\%, and performed a test to determine whether we rejected or accepted $\mathcal{H}_{0}$. 

We used the $T$-Student table with 28 degrees of freedom (total number of observations minus two) to determine the critical value. The critical value of t for a two-tailed test with 28 degrees of freedom and a significance level of 5\% is approximately 2.048. The observed t value was approximately -0.56. As the observed t-value (-0.56) is not greater than the critical value $\pm$ 2.048, we do not have sufficient evidence to reject $\mathcal{H}_{0}$. Therefore, we cannot conclude that there is a significant difference in the mean growth rate between male and female student enrollment over time at a significance level of 5\%. Thus, we accept $\mathcal{H}_{0}$. In this sense, it is essential to recognize the potential positive impact of the Meninas++ Project. The initiatives undertaken by Meninas++ contribute to the entry and retention of more women in STEM fields; however, external factors may mitigate these effects. Hence, it is important to continue expanding initiatives within municipalities and region to reinforce and extend the favorable outcomes of the project. 

\begin{figure}[!ht]
    \centering
    \begin{subfigure}{0.35\linewidth}
        \centering
        \includegraphics[width=\linewidth]{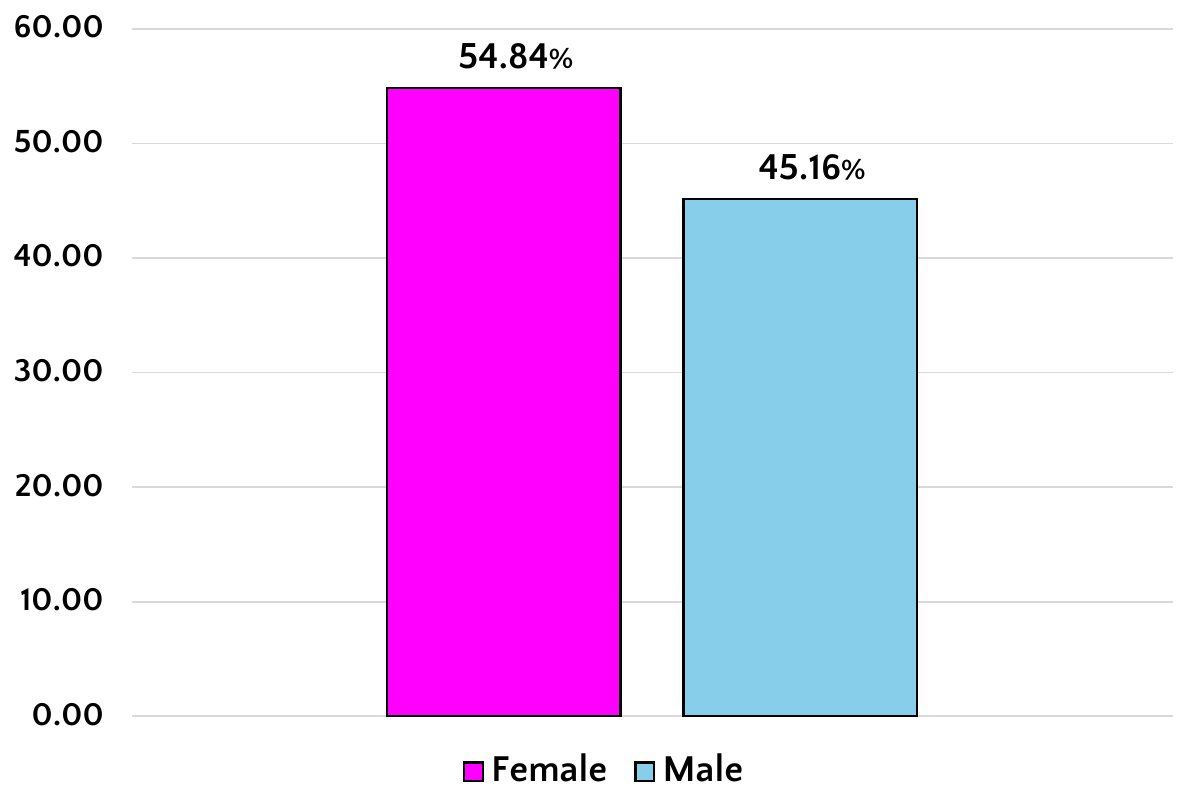}
        \caption{Gender.}
        \label{fig:chart1}
    \end{subfigure}
    \hfill
    \begin{subfigure}{0.3\linewidth}
        \centering
        \includegraphics[width=\linewidth]{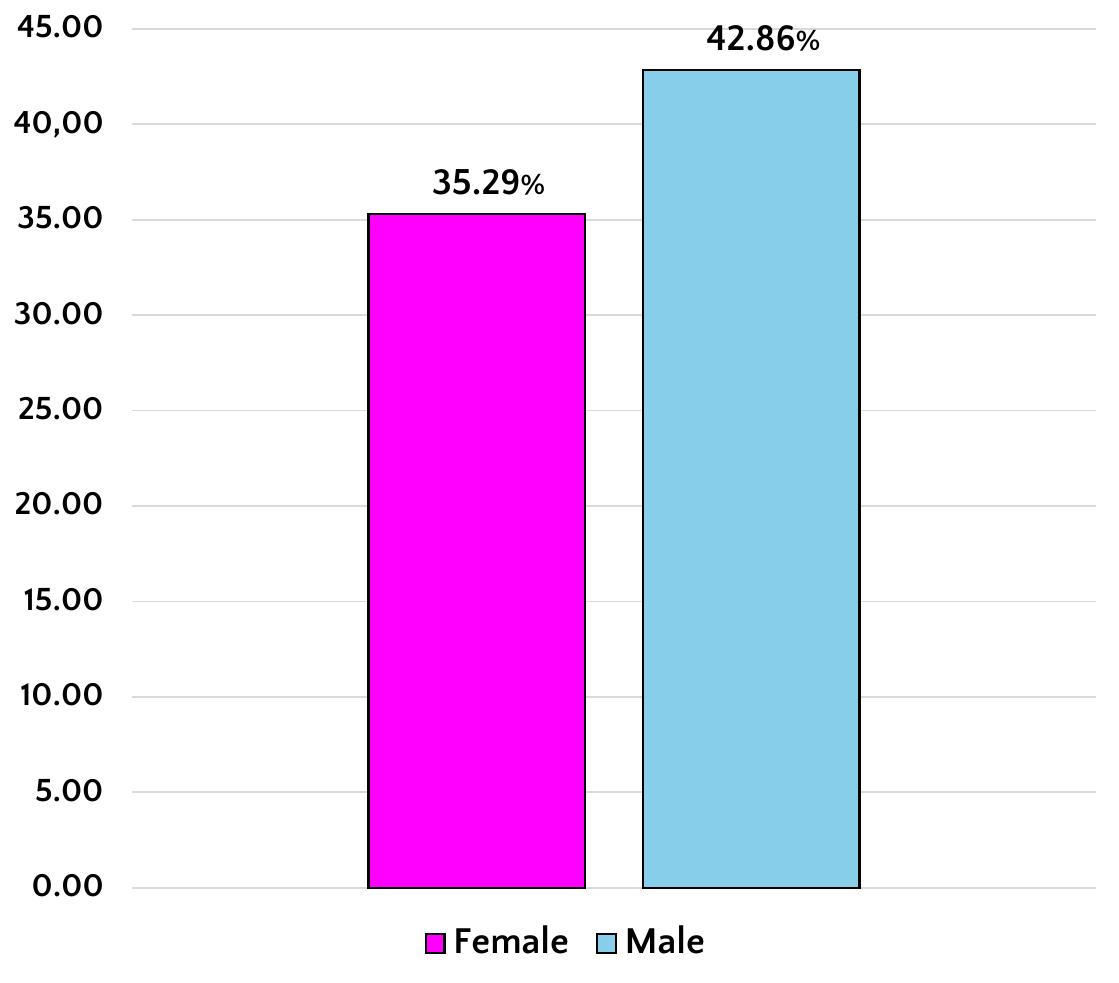}
        \caption{Exact Sciences.}
        \label{fig:chart2}
    \end{subfigure}
    \hfill
    \begin{subfigure}{0.32\linewidth}
        \centering
        \includegraphics[width=\linewidth]{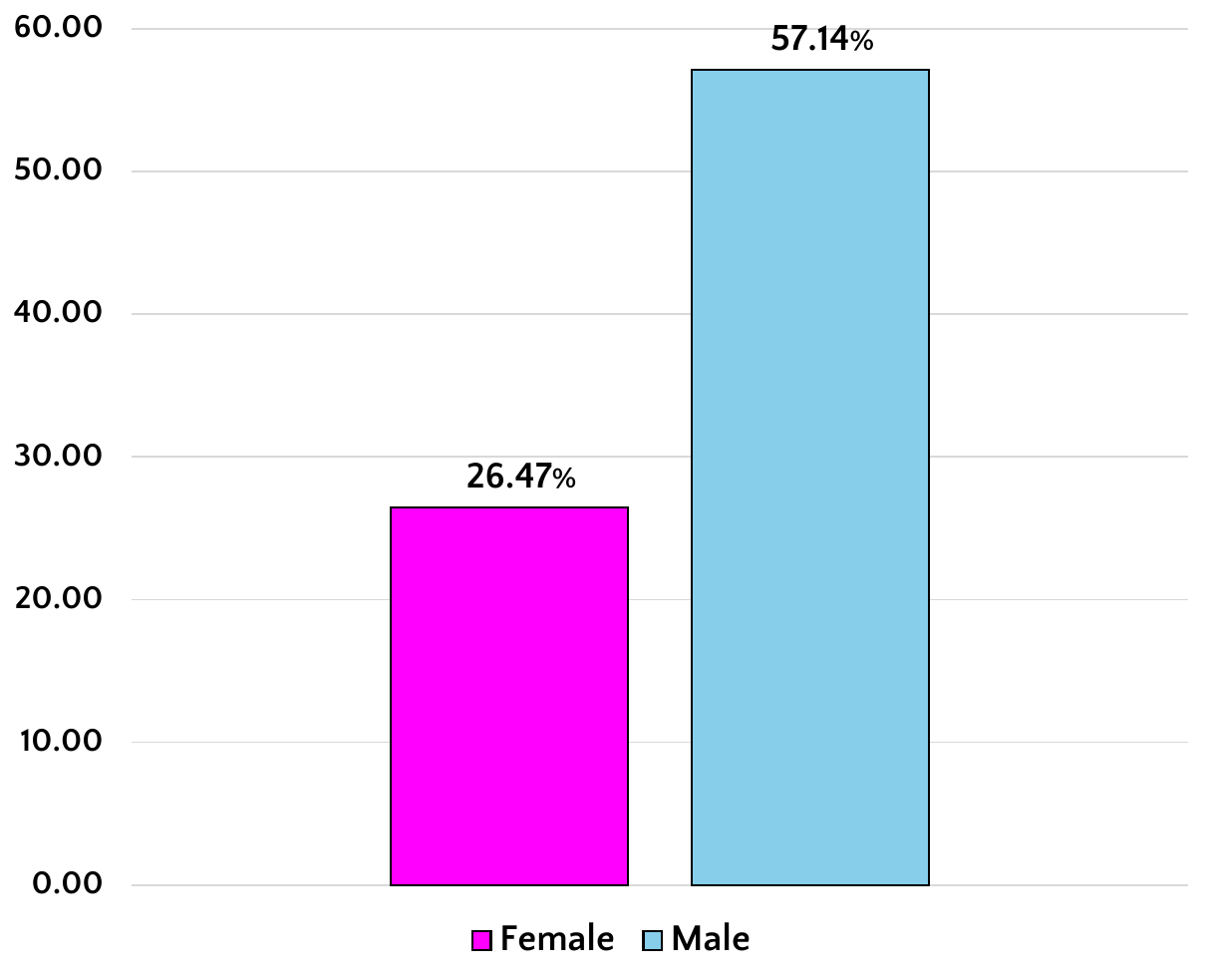}
        \caption{Computing.}
        \label{fig:chart3}
    \end{subfigure}
    \caption{Participant interests at the Ada Lovelace Day Community.}
    \label{fig:comparison}
\end{figure}

Annually, as part of our project, we host the Ada Lovelace Day event, an internationally recognized celebration. In 2023, we expanded our efforts by hosting two distinct events: ``Ada Lovelace Day Community'' and ``Ada Lovelace Day University'', which engaged both high school students from Rio Paranaíba and undergraduate students from the Systems Information Courses at UFV. The ``Ada Lovelace Day Community'' event was attended by 62 high school students and we conducted a survey aimed at discerning the profiles of high school students, exploring their perceptions of the technology field, and soliciting feedback about the event. Our results demonstrated that the majority of participants were female (54.84\%) between 17 and 18 years of age (Figure~\ref{fig:chart1}).

Our research aimed to evaluate students' interest in disciplines within the Exact Sciences, as depicted in Figure~\ref{fig:chart2}. The results demonstrated that 35.29\% of female students expressed interest in these fields, compared to 42.86\% of male students. While both genders showed interest, a higher percentage of male students were drawn to these fields compared to female students. As shown in Figure~\ref{fig:chart3}, we expanded our analysis to ascertain students' interest in the field of computing. The results suggests a significant gender gap: merely 26.47\% of female students expressed an interest in computing, in contrast to 57.14\% of male students. 
The proportion of male students interested by computing (57.14\%) exceeded that of those with a general interest in Exact Sciences (42.86\%). This disparity indicates a gender divide within the computing domain, suggesting that the allure of computing for male students transcends a mere inclination towards the Exact Sciences. However, the proportion of female students interested in computing was lower than that of those interested in Exact Sciences, declining from 35.29\% to 26.47\%. This decline in interest implies potential barriers for women in pursuing computing-related paths.

The ``Ada Lovelace Day University'' event saw participation from approximately 100 attendees, including 66 undergraduate students of Information Systems. The event featured a keynote titled ``Impacts of AI on Web Development'' delivered by a female Software Engineer from a major company. She is an alumna of the Meninas++ Project, holding a degree in Information Systems, and is currently active in the job market as a pleno. Her inspiring journey and involvement with the Meninas++ initiative served as a source of motivation for the attendees.

\begin{figure}[!htbp]
    \centering
    \begin{subfigure}{0.3\linewidth}
        \centering
        \includegraphics[width=\linewidth]{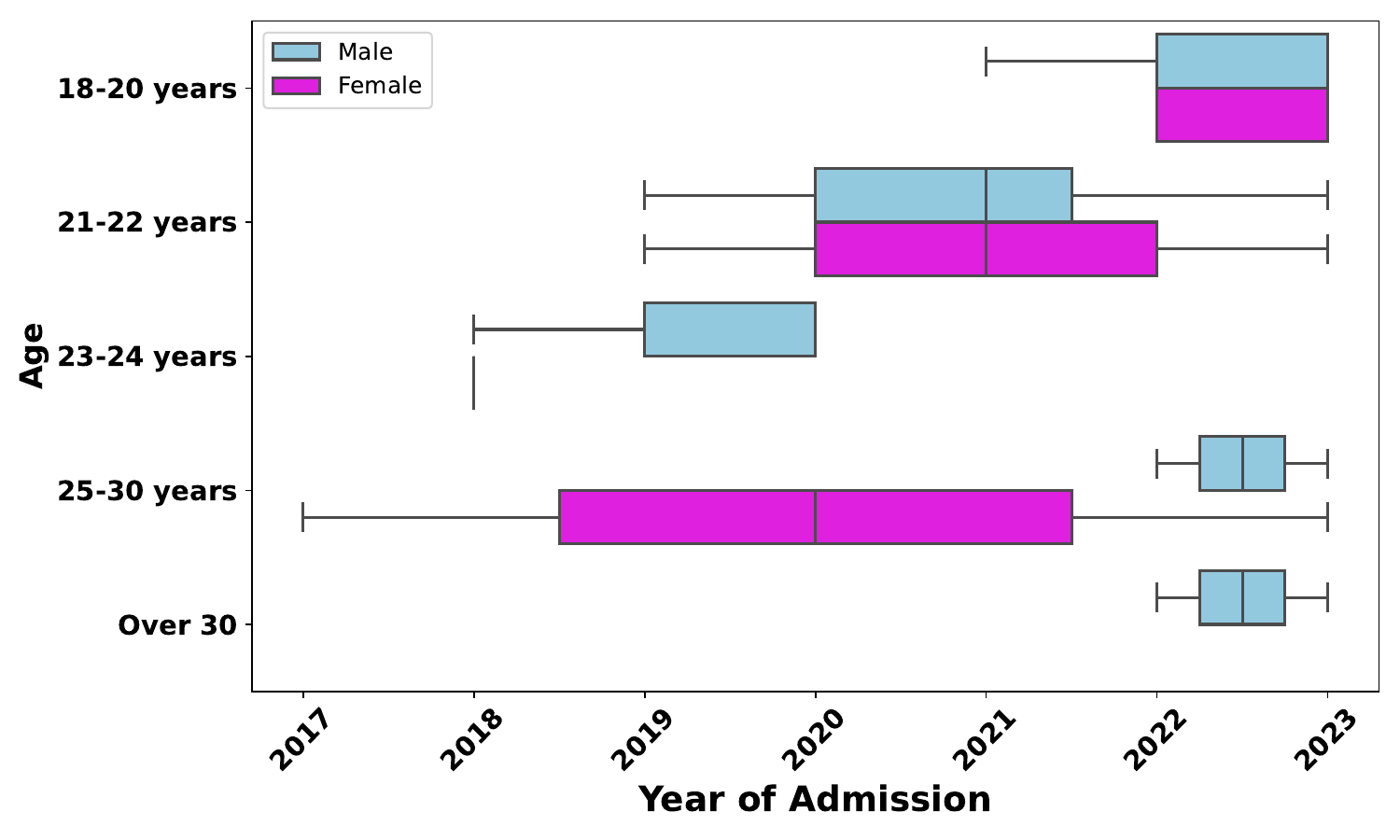}
        \caption{Age distribution by gender and year of admission.}
        \label{fig:age_by_gender_SI}
    \end{subfigure}
    \hfill
    \begin{subfigure}{0.37\linewidth}
        \centering
        \includegraphics[width=\linewidth]{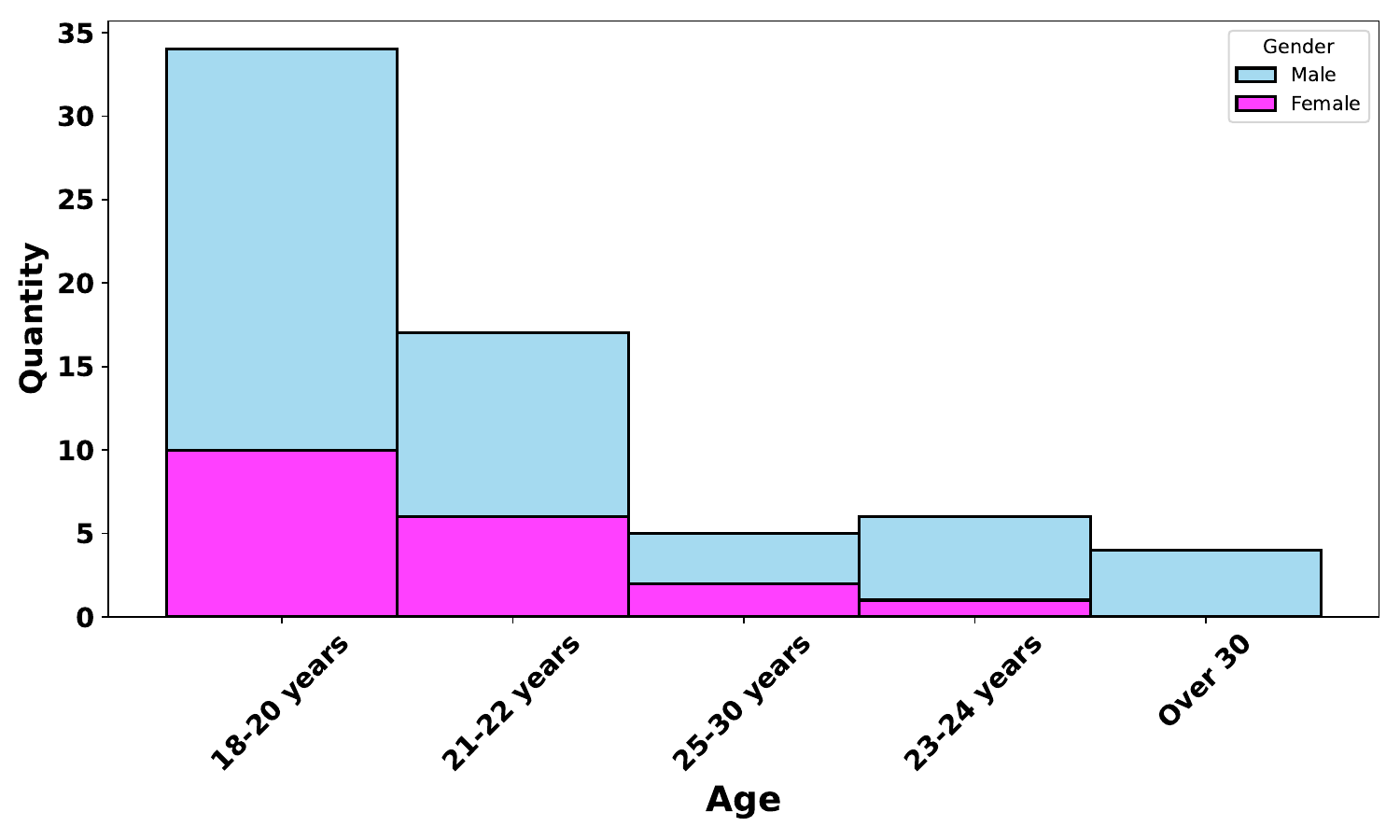}
        \caption{Age by gender.}
        \label{fig:histograma_idade_por_genero}
    \end{subfigure}
    \hfill
    \begin{subfigure}{0.3\linewidth}
        \centering
        \includegraphics[width=\linewidth]{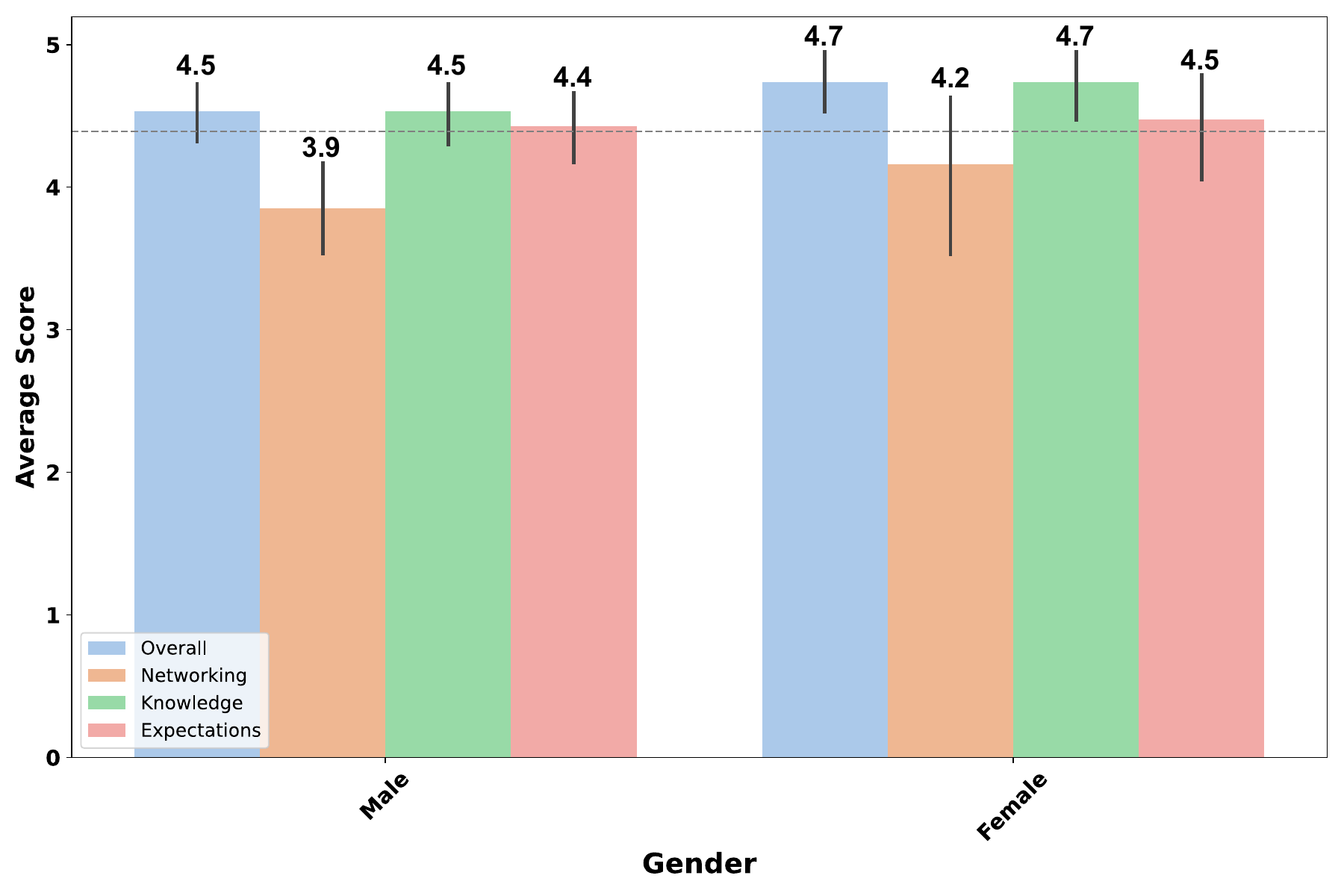}
        \caption{Evaluation scores by gender.}
        \label{fig:evaluation_by_gender}
    \end{subfigure}
    \caption{Participants of the Ada Lovelace Day University.}
    \label{fig:age-gender}
\end{figure}

First, we analyzed the age and year of university enrollment. Figure~\ref{fig:age_by_gender_SI} illustrates the dispersion of this information, allowing us to conclude that for both male and female students, there appeared to be no significant change in their age range over time. Additionally, we analyzed the distribution of age among male and female students, as depicted in Figure~\ref{fig:histograma_idade_por_genero}. 
We found that the majority of participants in the event were between 18-20 years old, primarily comprising students in the early stages of the Information Systems course. Our results indicate a significant level of engagement among male students in the event, demonstrating their support and commitment to the project's initiatives. This observation highlights the critical role that male students play in promoting gender diversity and equality in computing. The presence and endorsement of female and male students contributes to promoting gender equality efforts, thereby magnifying the project's influence.

We evaluated the event across various dimensions to gauge its effectiveness and impact on participants' experiences (Figure~\ref{fig:evaluation_by_gender}), including networking opportunities, acquisition of new knowledge, and alignment with participants' expectations. Ratings were measured on a scale ranging from 1 to 5, with 1 representing the lowest score and 5 the highest. The horizontal dashed line denotes the overall mean evaluation. Our results suggests that both male and female attendees had favorable experiences, suggesting that the event effectively met their expectations and provided valuable networking opportunities and knowledge acquisition, regardless of gender.

%Upon analysis, we observed that both male and female participants generally provided high evaluations across all categories. However, a slight disparity emerges in the average ratings between genders, particularly in aspects such as networking opportunities and acquisition of new knowledge, where females tend to rate slightly higher than males. This difference can be attributed to the event's primary focus on female empowerment and inclusion. The horizontal dashed line denotes the overall mean evaluation, irrespective of gender. Notably, most of the average ratings surpassed this benchmark, indicating that participants viewed the event and its activities positively, overall. These findings affirm that both male and female attendees had favorable experiences, suggesting that the event effectively met their expectations and provided valuable networking opportunities and knowledge acquisition, regardless of gender.

\begin{figure}[!ht]
  \centering
  \includegraphics[width=0.49\linewidth]{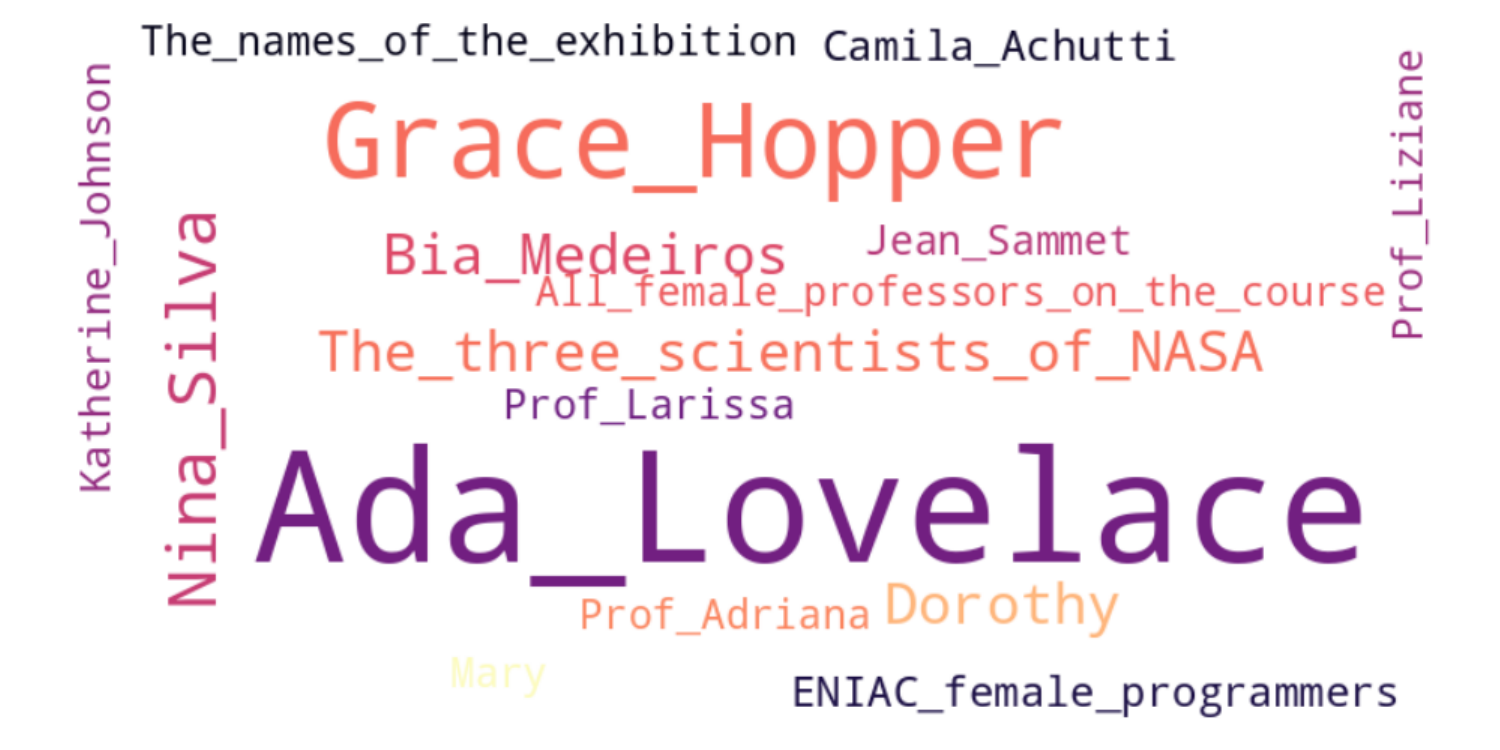}
  \caption{Names of influential women in Computing most cited by participants.}
  \label{fig:cloud-words}
\end{figure}

%Throughout the week of ``Ada Lovelace Day'', participants had the opportunity to visit the exhibition ``TechWomen: Women Who Transformed Technology.'' This exhibition attracted over 150 visitors, engaging students, educators, and professionals from various levels and disciplines, the exhibition not only showcased the achievements of women in technology, but also facilitated networking, mentorship, and knowledge-sharing opportunities across different segments of the academic and professional community.
During the ``Ada Lovelace Day University'', participants were asked to mention the names of women with a history of computing. Thus, we generated a word cloud to visualize the most frequently mentioned names (Figure~\ref{fig:cloud-words}). 
Our results indicated that many names mentioned by the participants were featured in the exhibition ``TechWomen: Women Who Transformed Technology''. We can observe the prominence of the names associated with pioneering women in computing and technology. Names such as ``Ada Lovelace,” ``Grace Hopper,” and ``Nina Silva'' stand out, reflecting their significant contribution to the field. Additionally, terms such as ``ENIAC female programmers'' and ``all female professors on the course'' further highlight the achievements of women in technology and the impact of the Meninas++ Project.

%Finally, within the scope of the Meninas++ Workshop, we orchestrated the Hackathon OMR Challenge, drawing on participation from 25 teams representing various regions across Brazil. Within these teams, women constituted 28\% of all competitors. Among these female participants, 13\% were part of Hackathon finalist teams. This outcome shows the talent, creativity, and dedication demonstrated by individuals of all genders in addressing challenges and driving innovation in the technological industry.

\section{Concluding Remarks}\label{sec:concluding_remark}

%This study sheds light on the actions undertaken by the Meninas++ Project to address the under-representation of women in STEM fields, particularly in computing. Through a comprehensive analysis, we demonstrated the significant impact of these actions in inspiring, empowering, and retaining female students in secondary and higher education. By leveraging both quantitative and qualitative data, we provided valuable insights into the effectiveness of our actions in fostering a supportive learning environment.

%Our findings contribute to the growing body of literature offering valuable insights for policymakers, educators, and stakeholders. By championing inclusion at every level of education and industry, we can create a future in which all individuals, regardless of gender, have equal opportunities to thrive and contribute to the ever-evolving landscape of computing and technology. 

This study sheds light on the actions undertaken by the Meninas++ Project to address the under-representation of women in STEM fields, particularly in computing. Our analysis revealed the impact of these actions in motivating, empowering, and retaining female students. By employing both quantitative and qualitative data, we provided meaningful insights into the efficacy of our initiatives in fostering a supportive learning environment.

In future work, we plan to conduct longitudinal studies to track the long-term impact of our project, deepen our analysis through qualitative research methods, explore intersectionality in STEM experiences, compare outcomes across institutions and regions, develop professional development programs for educators, foster industry partnerships, and analyze policies to support and create opportunities for women in the STEM field.

\section*{Acknowledgments}

%Omitted due to the double-blind review.
We would like to thank FUNARBE and CAPES for the financial support. This study was financed in part by the Coordenação de Aperfeiçoamento de Pessoal de Nível Superior - Brasil (CAPES) - Finance Code 001.

\bibliographystyle{sbc}
\bibliography{references}

\end{document}